\documentclass{elsarticle}
\usepackage{jasr}
\usepackage{framed,multirow}
\usepackage{amssymb}
\usepackage{latexsym}
\usepackage{graphicx}
\usepackage{natbib}
\usepackage{float}
\usepackage{xcolor}
\usepackage{anyfontsize}
\usepackage{parskip}
\usepackage{geometry}
\usepackage{url}

\journal{Advances in Space Research}

\begin{document}


\begin{frontmatter}

\title{Data Pipeline Architecture and Development for VELC onboard Space Solar Mission AdityaL1}

\author{Jagdev Singh \corref{cor1}} 
\cortext[cor1]{Corresponding author: 
  email: jsingh$@$iiap.res.in;}
\author{B. Raghavendra Prasad \fnref{fn1}}
\fntext[fn1]{Principal Investigator-VELC,AdityaL1(ISRO)}
\author{ Chavali Sumana, Amit Kumar, Varun Kumar, Muthu Priyal, Suresh Venkata}

\address{Indian Institute of Astrophysics, Koramangala, Bengaluru - 560034}

\begin{abstract}

	ADITYA L-1 is India's first dedicated mission to study Sun and its atmosphere with Visible Emission Line Coronagraph (VELC), a major payload on ADITYA-L1. VELC has provision to make imaging and spectroscopic observations of the corona, simultaneously. Imaging with the Field of View (FOV) from 1.05Ro to 3Ro will be done in continuum at 500nm. The spectroscopic observations of solar corona in three emission lines, namely 5303 {\AA} [Fe XIV], 7892 {\AA} [Fe XI], 10747 {\AA} [Fe XIII], and Spectro-polarimetry at 10747 {\AA} [Fe XIII] will be performed with FOV of 1.05-1.5Ro. In this work, the end-to-end data pipeline architecture and development of the VELC payload are presented. The VELC proposal submission form, satellite observation parameters, data products, level definitions, data pipeline and analysis software to process the big raw data sets obtained using VELC instruments onboard satellite to science-ready data are discussed.

\end{abstract}


\end{frontmatter}

\section{Introduction}

To understand the physical and dynamic characteristics of the solar corona, observations were made during the total solar eclipses, where the moon acts as a natural occultor to block the solar disk. In 1930 Lyot invented coronagraph by blocking the solar disk, enabling the observation of the extended coronal atmosphere of the Sun. This invention opened the doors to study solar corona in the times other than a total solar eclipse, with ground base coronagraphs as well as space based instruments. Number of space mission such as SOHO, SDO, STEREO, Trace, Ulysses, Hinode, Solar Orbitor,  Parker Solar Probe and many others have made observations of the sun and its atmosphere in different wavelength domain such as X-rays, EUV and UV mostly. Some are still in operation. 
ADITYA –L1 is the first Indian solar mission to be launched in the L1 orbit of the Sun earth system to study the sun and solar atmosphere. The Visible Emission Line Coronagraph (VELC) (\citet{singh2011} and \citet{prasad2017}) will be one of the major instruments onboard ADITYA-L1. VELC is an internally occulted coronagraph designed to study the solar corona. The coronagraph is capable of recording the solar corona from 1.05Ro to 3Ro in continuum at 5000 {\AA} with an image scale of 2.51 arcsec $/$ pixel and a cadence of 30 seconds. VELC has the facility of multi-slit spectroscopy at three emission lines 5303 {\AA} [Fe XIV], 7892 {\AA} [Fe XI] and 10747 {\AA} [Fe XIII] with a spectral resolution of 28 m{\AA}$/$pixel, 31 m{\AA}$/$pixel and 202 m{\AA}$/$pixel, respectively in FOV 1.05Ro to 1.5Ro with an image scale of 1.25 arcsec$/$pixel. VELC also has a dual-beam spectro-polarimetry channel for magnetic field measurements at 10747 {\AA} (\citet{prasad2017} and \citet{kumar2018}). The payload has four detectors (three sCMOS  and one InGaAs ) to enable imaging, spectroscopic and spectro-polarimetric  observations simultaneously. One for imaging the solar corona in continuum and other two for recording spectra in two visible emission lines 5303 {\AA} and 7892 {\AA}. InGaAs detector is used for the spectroscopic and spectro-polarimetric observations in 10747 {\AA}. The science  objectives  of  VELC are (\citet{singh2013}) (i)diagnostics  of  the corona  and  coronal  loops  plasma  (temperature, velocity  and  density)$;$  (ii)heating  of  the  corona$;$  (iii) development, dynamics  and  origin  of coronal  mass  ejections  (CMEs)$;$  (iv) studies  on  the  drivers  for  space weather and (v) measurement  of  coronal magnetic fields. More details can be seen is \citet{singh2019}. In this article, the VELC proposal submission form, satellite observation parameters, data products, level definitions, data pipeline and analysis software to process the large volume of data obtained using VELC instruments onboard satellite are discussed.

\section{VELC Data Pipeline Development}
\begin{figure}[ht]
\begin{center}
\includegraphics[width=3.4in, scale=0.5]{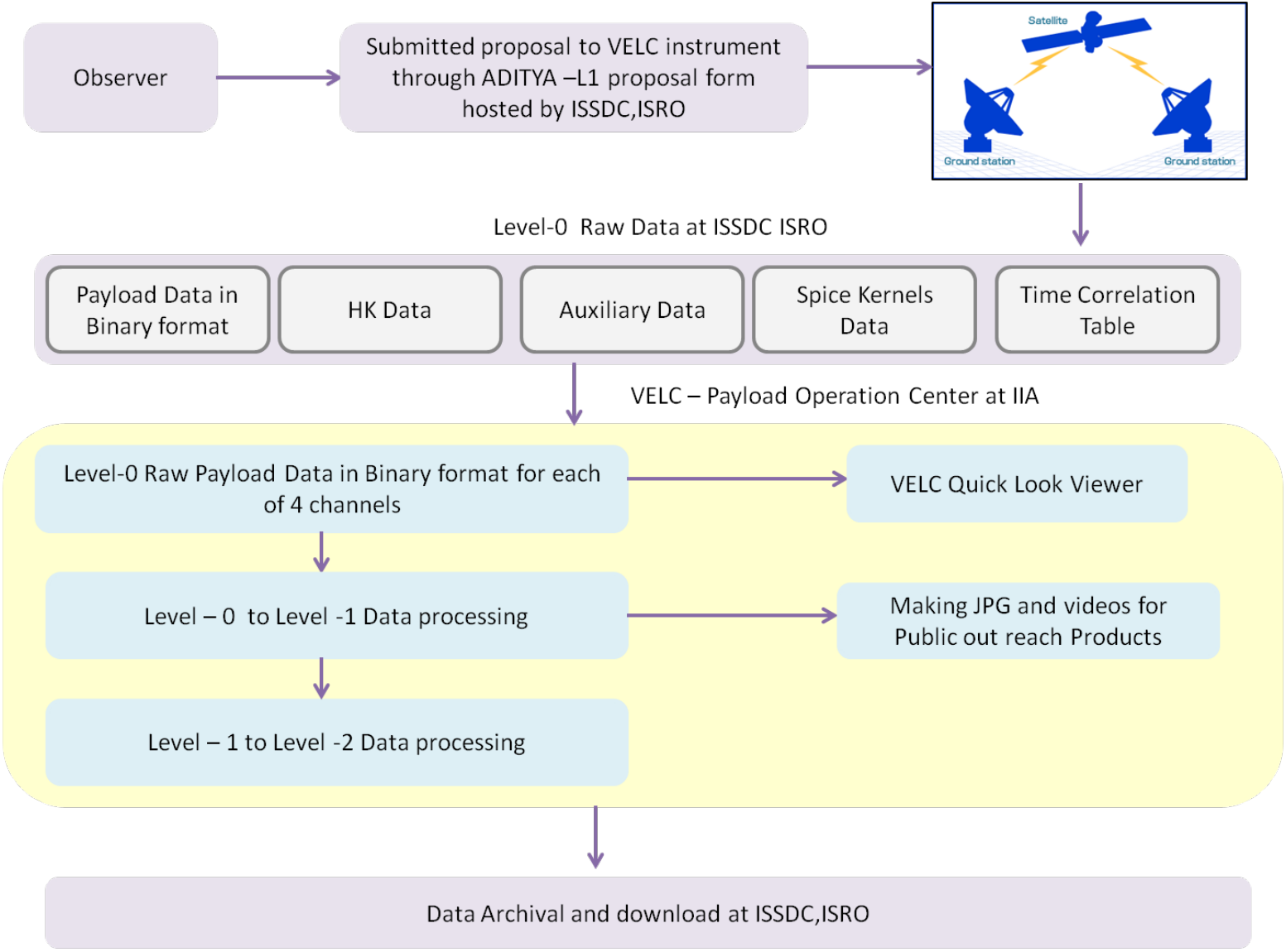}
\caption{Flow chart of VELC end-end data pipeline architecture.}
\label{fig:1}
\end{center}
\end{figure}

After the satellite has been stabilised in the L1 orbit, the operation to test the functions of various components will begin online when the satellite is in line-of-sight. After that observations need to be made to test data pipeline and its various subroutines such as proposal submission form, observation sequence, data saved onboard and download at ground, format of the data and finally confirm the calibration of the payload. In this section, the end-to-end Data pipeline Architecture for VELC, as shown in Figure \ref{fig:1} is discussed.

After the Performance Verification (PV) phase, various procedures to obtain and analyze the data will be reconfirmed during the regular observations made during Guaranteed Time (GT) phase and put in the public domain. Meetings and workshops will be conducted to share all the procedures with the interested scientists. Some synoptic observations will be carried on daily basis with flexible parameters. In addition, the observational proposals will be invited from the scientists all over the world through a web-based portal in response to ISRO's advertisement.  List of selected proposal by a committee on scientific merit and feasibility will be made and observational schedule prepared on daily basis considering the data volume that can be downloaded per day. Then the observing plan will be up loaded for execution on daily basis.

Post observations, the data obtained using all the four detectors and saved in the onboard memory will be downloaded at ISRO Telemetry, Tracking and Command Network (ISTRAC), Bengaluru as and when satellite comes in the line-of-sight of the antenna. The download of data will be done by Indian Space Science Data Centre (ISSDC) at ISTRAC. Various data processing levels like Level-0(L0), Level-1(L1) and Level-2(L2) have been defined for the VELC data. Raw data files are inputs to L0 to generate the files in binary format. The compresses data will be decompressed at ISSDC to generate L0 data. The output L0 files, Spacecraft Planet Instrument C-matrix Events (SPICE) kernels, Time-tag correlation tables, Telemetry House Keeping (HK) data, and logs files will then be downloaded at Payload Operation Center (POC). The L0 files are inputs to generate L1 data and will be used to create offline (Quick Look Display) QLD. Similarly, L2 will be developed at VELC-POC, corrected for the instrumental effects and with proper observations information in the header. All these data files, L0, L1 and L2 will be sent back to ISSDC for data archival. Further, processing of the data will be done as per the scientific requirements. In the following the procedures are described in detail.

\section{VELC proposal submission form}

VELC proposal submission form is a user-friendly web-based portal for observers to submit their observation plan for VELC payload.

\subsection{ \textbf{VELC modes of operation}}
There are 4 channels, one for each detector. The details for each channel are given below. 
\subsubsection{\textbf{Continuum channel}}
In the continuum channel images the solar corona can be obtained continuously from 1.05Ro to 3Ro (Figure \ref{fig:2}a) at 5000 {\AA} ±5 {\AA} with a pixel sampling of 2.5, 5 and 10 arcsec/pixel by binning the data onboard. It has three modes of operation: 1) Synoptic mode option-1, 2) Synoptic mode option-2, 3) Continuum proposal mode.

\begin{description}
\item [Synoptic mode option -1]
In this mode of operation, analysis of the images is done onboard continuously, using hardware developed at Space Applications Centre (SAC), Ahmedabad to detect the occurrence of Coronal Mass Ejection (CME). Once occurrence of CME is confirmed onboard, the data saving mode becomes active and images of the solar corona will be saved with a cadence of 1/2/4/5 min (cadence will be decided based on the PV phase observations). During the period when no CME is detected, the images will be saved at an interval of 3 Hrs. This methodology was adopted to obtain the data with relatively high spatial resolution and manage the volume of data within limits (about 30 Gbits). This methodology has the limitation of not having the data before and after the event, which is very much needed to study trigger mechanism of CME and physical and dynamical characteristics immediately before and after the energetic events. Further, there is possibility that saving of the images stopped before the CME moves out the FOV. It may be noted that in this mode exposure remains fixed by the hardware and cannot be varied.

\item [Synoptic mode option -2]
Alternately, images of the solar corona without using the CME detection hardware, with certain cadence and pixel resolution for 24 hrs a day and save all the images onboard without exceeding the volume of the data limit can be taken. This is possible by optimising the spatial and temporal resolution of the images. The possible combinations can be made by choosing cadence of 1, 2, 4 or 5 minuets and pixel binning 1 x 1, 2 x 2 or 4 x 4. It is also possible to choose different exposure time for the data set. For example, with pixel binning of 2 x 2 and cadence of 2 minutes, the volume of data for 24 hours a day will be about 16 Gbit with data compression factor of 2. Various combinations of exposure time and pixel binning can be done to obtain the images for 24 hrs a day within limited data volume. Further, this methodology has the advantages of having the data before and after all the energetic events. Immediately, after the download, all the images will be subjected to software code to detect the occurrences of CME or energetic event in the solar corona. The Table 1 lists the comparison between the two methods. 

\item [Continuum proposal mode]
It may be noted that if required for some scientific objective, it is possible to take the coronal images at fast cadence of about 10 seconds but for short time of about 45 minutes, as the volume of data will exceed the limits to save and download.

\end{description}

\begin{table}[ht!]
\begin{center}
\caption{Comparison of synoptic mode option-1 and synoptic mode option-2}
\begin{tabular}{ |p{6cm}|p{6cm}| }
 \hline
\textbf{Synoptic mode option-1}  & \textbf{Synoptic mode option-2}  \\
 \hline
 
 Relatively high spatial and temporal resolution & Relatively low spatial and temporal resolution \\
\hline
Images available 2 minutes after the occurrence of CME  & Images available before and after  the occurrence of CME  \\
\hline
 Possibility to stop the data saving before the CME moves out of FOV & All the images will be saved irrespective of CME is there or not in FOV \\
\hline
 Only active coronal studies will be possible & Quiet coronal studies are also possible such as streamer structures, polar regions with time \\
\hline
 Knowledge of physical and dynamical properties of the corona before and after the CME will not be available which is, an important factor & Knowledge of physical and dynamical properties of the corona before and after the CME will be available which is, an important factor \\
\hline
Possibility of missing and false detection of CME & Not applicable as all the data is downloaded \\
\hline
Information about occurrence of CME as soon as data is downloaded  & Information about occurrence of CME after analysis of downloaded data,  $\approx$ 12 hrs delay \\
\hline
\end{tabular}
\end{center}

\label{tab:1}
\end{table}

\subsubsection{ \textbf{Spectroscopy channels}}
\begin{figure}[ht!]
\centerline{\includegraphics[width=3.4in, height=3.2in,scale=0.5,keepaspectratio]{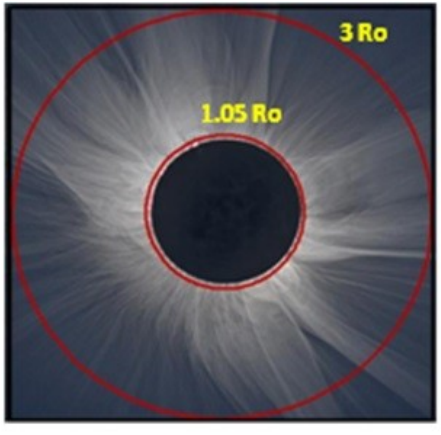}\includegraphics[width=3.4in, height=3.5in,scale=0.5,keepaspectratio]{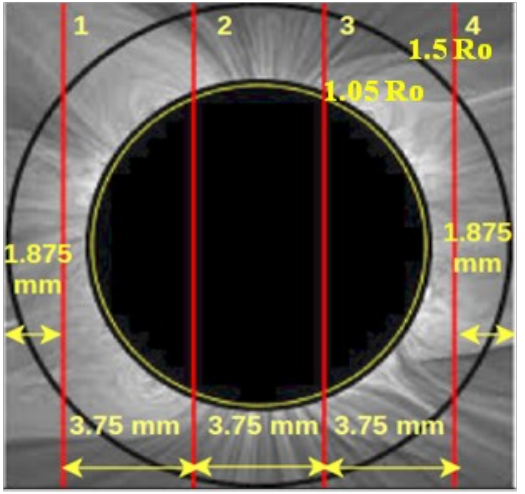}}
\caption{Figure 2a shows the FOV in red circles (up to 3 Ro) of continuum channel of VELC with the processed solar coronal image take during total solar eclipse of 2010. Figure \ref{fig:2}b shows the FOV by black circles (up to 1.5 Ro) of spectroscopic channels of VELC with the processed solar coronal image take during total solar eclipse of 2010 (\citet{samanta2016}). The figure shows the location of the coronal image on 4 slits of the spectrograph with LSM at home position.}
\label{fig:2}
\end{figure}

 The multi-slit spectrograph is designed to study the solar corona in three emission lines 5303 {\AA} [Fe XIV], 7892 {\AA} [Fe XI], and 10747 {\AA} [Fe XIII] with FOV 1.05Ro to 1.5Ro.  The plate scale on the plane of the slits is 192.3 arcsec/mm. Thus, the width of each slit is 50 microns equivalent to 9.6 arcsec at the multi-slit plane. At the spectral  domain, a 6.5 micron pixel size of CMOS detector corresponds to 1.25 arcsec for the 5303 {\AA}   and 7892 {\AA}   emission lines. But, the 25 micron pixel size of the IR camera corresponds 4.8 arcsec. It has four modes of operation, namely 1) Sit and stare 2) Raster scan 3) IR Spectro-polarimetry mode 4) Spectroscopic CME mode.
    \begin{enumerate}
    \item \textbf{Sit and stare mode}
     The coronal image is focused on 4 slits of the spectrograph at pre-programmed location. The spectral images are obtained at the fixed locations on the solar corona, with required exposure times (between 200 ms to 50 sec) and cadence by the observer. The observer can choose these from the set of parameters available in the proposal form. Exposure times will be optimized post in-orbit performance verification phase. Other parameters such as pixel binning, setting of camera gain etc. can be selected by the observer.

\begin{table}[ht!]
\begin{center}
\caption{VELC Modes of operations, respective exposure times and cadence range.}
\begin{tabular}{ |c|c|c| }
 \hline
\textbf{Mode of operation} & \textbf{Exposure time range} & \textbf{Cadence range}   \\
 \hline
 
 Continuum proposal mode & 150 ms to 50s in certain steps & 30s to 300s \\
 
\hline
Visible line Spectroscopy sit and stare mode & 200ms to 50s in certain steps & 2s to 600s \\

\hline
 Visible line Spectroscopy Raster Scan mode & 1s to 50s in steps & 1s to 600s \\
 
\hline
 IR Spectroscopy sit and stare mode & 103ms  to 50s & 1s to 600s  \\
\hline
IR Spectroscopy Raster Scan mode & 103ms to 50s & 1s to 600s \\
\hline
IR Spectro polarimetry mode & 106 ms & 625ms \\
\hline

\end{tabular}
\end{center}

\label{tab:2}
\end{table}

\item \textbf{Raster scan mode}
     Various parameters to make observations in this mode can be chosen as in case of sit and stare mode. In addition, one needs to choose the parameters of Linear Scan Mechanism (LSM) to move the image on the multi-slit of the spectrograph in steps. The four slits (50 micron each) are separated by 3.75 mm as shown in Figure 2b. To make the full image of the solar corona in emission line up to 1.5 Ro, one needs to move the image on slits in steps and record the spectra at each location. In other words, make observations in raster scan mode. To cover the whole FOV, the image needs to move about 3.8 mm on the slits. Provision is made to move the image by 4 mm to have an overlap of data for 0.25 mm. Normally, the LSM is positioned at middle location, called home position. In this position, image of solar corona on the 4 slits will appear as seen in Figure \ref{fig:2}b. To make a raster scan to cover full FOV, one may move the LSM to -1.0 mm position and then move up to $+$1.0 mm position in steps of multiple of 10 microns. It may be noted that when LSM moves by 1 mm, the image moves by 2 mm. It is possible to plan the observations on part of the solar corona. Also, one can plan to add number of images to avoid the saturation of a single image and increase the signal to noise ratio of the data in both, sit and stare, and raster scan mode operation. More details of the parameters to make the observations are given in Table 2.

It may be noted that term cadence refers to time interval between the two spectra, not the interval between two raster scans. The interval between two raster scans depends on exposure time and number of steps chosen in case of multiple scans obtained. The 600 second cadence limit is kept in view various commands to take and save the spectra. The cadence of 625 ms for IR spectro- polarimetry is fixed to limit the average of 5 spectra within 22.5 degrees. More details about the spectro-polarimetry may be seen in (\citet{venkata2017};\citet{venkata2021}). We have also added the related references.

\item \textbf{IR Spectro-polarimetry mode}
      All the parameters of the camera and exposure time are fixed, not changeable. The coronal image will be positioned at a pre-programmed position with respect to slits. The retarder starts rotating continuously when one begins the observations, completing one rotation in 5.95 seconds. The 16 spectra with an interval of 22.5 degrees will be obtained per rotation. It can also be operated in raster scan mode. 

    \item \textbf{Spectroscopic CME mode}
     In this mode of operations, spectroscopic observations have been planned by fixing the parameters such as location of the image, exposure time and others. When CME is detected using the coronal images in continuum channel and choice is made to start spectroscopic CME mode operation in the proposal form, then spectroscopy channels can abort the running program and start observing in this mode.

\end{enumerate}

\section{Some details about the payload and observational parameters}
Various terms are used to define the instrument (payload) and observational parameters.  These terms are explained in the following: 
\begin {description}
\item [Exposure Time]
 The exposure time is the time to take a single image.
\item [Frame Time]
 If the user chooses to add certain number of images, with chosen exposure time, then those number of images are added to form a single image in the camera electronics hardware.  The total time to obtain that image is called ``Frame Time''.  
\item [Cadence]
 User may choose larger time interval than the exposure time or frame time, between two images to be saved in the onboard memory and later to download. This time interval between the two frames is generally called ``Cadence''. 

Note$:$ Based on the calibration data on board during the Performance Verification and Guaranteed Time phase, exposure times will be optimized and the list of suitable exposure times made to obtain good data and achieve the science goals

\item [Snapshot]
 Snapshot refers to save the images at certain interval. It is similar to cadence.
\item [Frame binning]
 Frame binning refers to adding number of images to make one image and save it in onboard memory to be downloaded when possible. 
\item [Spatial binning]
 Spatial binning is used to reduce the image's size by adding the intensity values of neighbouring pixels. Thus more observations can be made in view of the limit on data volume. VELC has the following onboard spatial binning options.
\begin{enumerate}
\item \textbf{Continuum channel options} – 1x1, 2x2, 4x4, 8x8$;$ Provision for 8x8 was kept for certain programs. Care should be taken as it can lead to saturation of the data (16-bit).
\item \textbf{Spectroscopy channels options} - 1x1, 2x1, 4x1, 8x1. The binning is possible in the spatial direction only to maintain the spectral resolution. 
\end{enumerate}

\item [Window $/$ ROI]
 This refers to a option where one can select a certain area of the image in continuum and certain portion of the spectra to reduce data volume.  In case of continuum channel and 3 spectroscopic channels of VELC, each has 4 defined windows and user can choose any one of those window options. Table 3 list the size and location of the windows on the detectors in pixels. Actually it is area on the detector for which data will be saved in the onboard memory depending on the chosen windows for each detector.

\begin{table}[ht!]
\begin{center}
\caption{VELC Window$/$ROI size and pixel locations in bracket for continuum channel and three spectroscopy channels}
\scalebox{0.8}{%
\begin{tabular}{ |c|c|c|c| }
 \hline
\textbf{Window Name} & \textbf{Continuum channel ROI} & \textbf{5303 {\AA} and 7892  {\AA} spectroscopy channels ROI} & \textbf{IR spectroscopy channel  ROI} \\
 \hline
Full Window & 2192 x 2592 (0:2191, 0:2591) & 2192  x 2592 (0:2191, 0:2591) & 512 x 640 (0:511, 0:639)  \\
\hline
 Left Window & 1096 x 2592 (0:1095, 0:2591)  & 1096 x 2592 (0:1095, 0:2591)  & 112 x 640 (400:511, 0:639) \\
\hline
 Right Window & 1096 x 2592 (1096:2191, 0:2591) & 1096 x 2592 (1096:2191, 0:2591) & 112 x 640 (0:111, 0:639) \\
\hline
 Top Window & 2192 x 1296  (0:2191, 1296:2591) & 2192 x 1296 (0:2191, 1296:2591) & 512 x 176 (0:511, 464:639)  \\
\hline
Bottom Window & 2192 x 1296 (0:2191, 0:1295) & 2192 x 1296 (0:2191, 0:1295)  & 512 x 176 (0:511, 0:175) \\
\hline
\end{tabular}}
\end{center}
\label{tab:3}
\end{table}

\item [Alternate gain frames]
 If one requires images to be saved with low and high gain alternatively, it is possible. This implies that first one saves the image with low gain only and discard the high gain image and for next image, one discards low gain image but saves high gain image.  But the exposure time will be same for both the images. 
 
 \item [Gain]

 \begin{itemize}
    \item CMOS detector used in continuum and two spectroscopic channels has the capability to provide data both high and low gain, simultaneously, with 11-bit format. 
    \item There are two low gains namely 1x and 2x and two high gains 10x and 30x. 
    \item User needs to choose one low gain and one high gain out of four available. Thus 4 combinations of gains are (1x and 10x),(1x and 30x),(2x and 10x) and (2x and 30x).
    \item For the IR channel there are two gain, low gain and high gain. The user needs to select one of the two gains. The data format is 12-bit.
\end{itemize}
\item [Payload configuration]
VELC is internally occulted coronagraph with an entrance aperture of 145 mm and reflective objective of 195 mm. There is a mechanical shutter before the entrance aperture which can be opened or closed whenever required. The shutter is fitted with a 45 mm diameter reflecting type neutral density filter of optical density 4 to limit the sunlight entering the coronagraph when in closed position. Lyot- stop removes the diffracted sunlight from the edges of the entrance aperture more details may be seen in (\citet{prasad2017}). We have kept all the observational parameters such as exposure time, spatial binning, image binning, gain setting etc., very flexible in view of the unknown amount of scattered light in the instrument after the launch. Care is being taken to minimize the contamination before and during the launch that is likely to increase scattering of disk sunlight from the primary mirror. These parameters will be optimized after the launch and will be shared with all the scientists.

\item Multi-slit and LSM details
\begin {enumerate}
    \item Number of slits$:$ 4 fixed slits.
    \item Slit width$:$ 50 \textmu m 
    \item Slit height$:$ 15mm
    \item Slit Separation $:$ 3.75mm (From centre of one slit to centre of adjacent slit)
   \item Linear Scan Mechanism (LSM) has two mirror mounted on a device so that both the mirrors (together) move in steps such that it shifts the image on the slits of spectrograph by 2 times the distance it moves. For example, it the mirrors move by 1 mm, the image shifts by 2 mm. The step size can be chosen in multiple of 10 microns. The time taken for a step of 10 microns is 0.8 second. Table 4 lists the parameters for the LSM need to be chosen for the raster scan observations.  
\end{enumerate}

\begin{table}[ht!]
\begin{center}
\caption{Different parameters for raster scan observations to be defined.}
\scalebox{0.8}{%
\begin{tabular}{ |c|c| }
 \hline
\textbf{ Starting position} & \textbf{User can define any position between -1 to $+$1 mm} \\
 \hline
 Movement direction  & User can choose the direction of the movement of the LSM to scan the FOV during the observation.\\
\hline
Step size & This step-size is the distance LSM moves to the next position and stay there until it gets the next command \\
\hline
Number of steps & User defines the number of steps for which LSM moves by a defined step-size required for the observation \\
 \hline
 Time interval between steps & User defines the time after which the LSM becomes active to move mirrors from one step to another step \\
\hline
\end{tabular}}
\end{center}
\label{tab:4}
\end{table}
\end{description}

\section{Onboard Calibration Plans}
\begin {description}
\item [Dark Calibration]
 The mechanical shutter just before the entrance aperture has a 45 mm reflecting type neutral density (ND) filter with density 4 which permits part of the sunlight to enter the coronagraph when pointed towards the sun. Therefore, satellite is planned to be pointed 15 to 30 degree away from the sun direction so that no light enters the VELC to confirm the dark current value onboard with that of laboratory value. Then images will be obtained for all the required exposure times using all the detectors with the mechanical shutter closed. The dark images will be compared with those obtained in the laboratory before the launch. While pointing the satellite for this purpose, care will be taken that no bright source (star or celestial object) is within the FOV.

\item [Disk Image and Spectroscopic Calibration]
 The ND filter mounted at the center of the mechanical shutter permits part of sunlight to enter the coronagraph when pointed towards the sun or off up to about 1.5 degree. If the satellite is pointed about 32 arcmin away from the sun in any direction then the full image of the solar disk will be made on the CMOS detector in continuum channel. Thus, by off pointing satellite 32 arcmin away from central position of the sun in the four directions, north, south, east, and west solar disk images will be take. Figure 3 shows the FOV of the VELC in continuum channel and composite pictures of disc images obtained in the 4 directions. If the satellite is off pointed by 12 arcmin in the vertical or horizontal direction, the solar disk light will fall on two the slits. Therefore, by off pointing the satellite by 12 arcmin in the four directions we will be able to take spectra of the solar disk with all the slits. The transmission of the filter has been calibrated at the required wavelengths to compute the exact part of the sunlight entering the instrument at those wavelengths. The smaller aperture and 4 density of the ND filter cuts the sunlight by a factor of 7x10$^{-6}$ making the brightness of disk image similar to that of solar corona. Therefore, we shall be able to take the solar disk (shutter closed) and coronal images (shutter open) with similar exposure times.

\end {description}

\begin{figure}[h]
\centerline{\includegraphics[width=3.4in, scale=0.5]{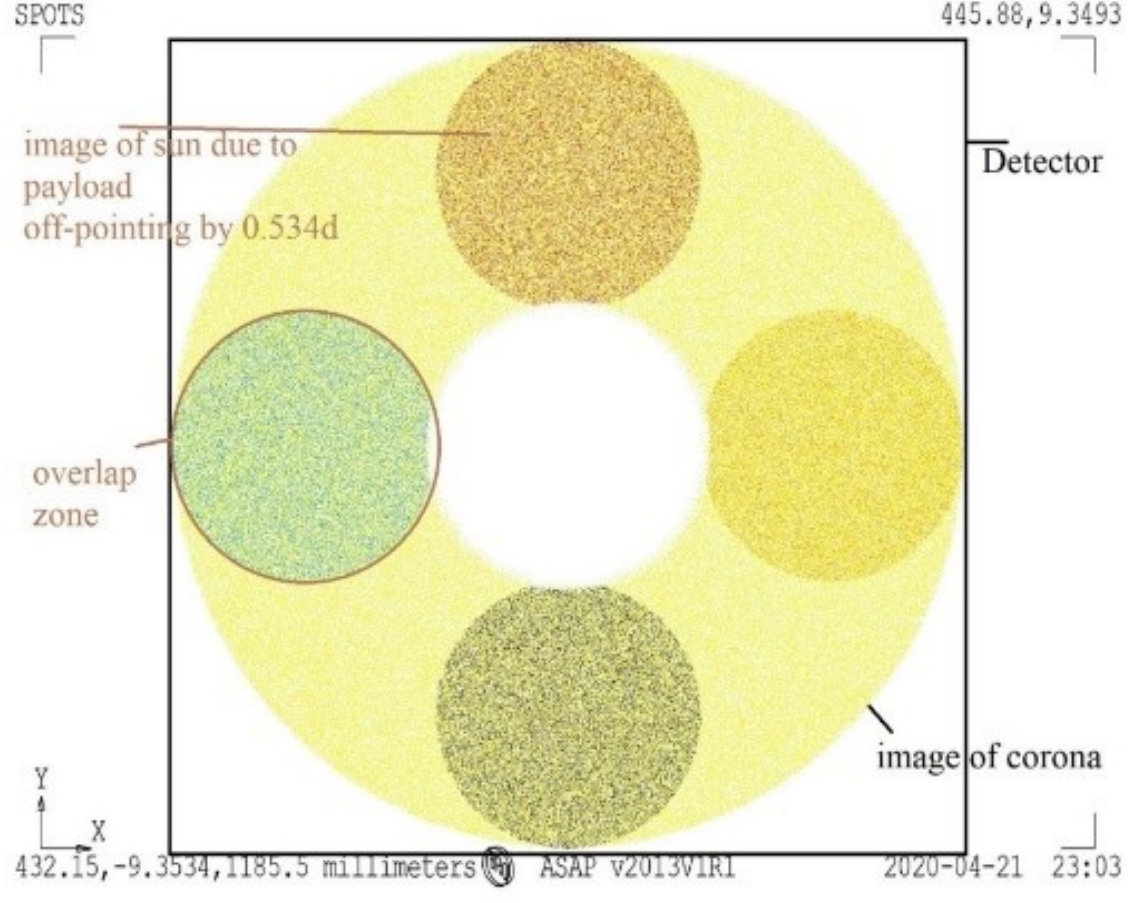}}
\caption{The rectangular portion marked by dark lines shows the size of the detector. The yellow ring depicts the area of coronal image in continuum channel and white portion the sun blocked by the occultor when satellite is centred on the sun. The 4 circular areas in different colours indicate the disc images when satellite is pointing ± 32 arcmin away from the sun in north, south, east and west. This is composite figure of all the 4 directions.}
\label{fig:3}
\end{figure}

\section{VELC data products and data level definitions}

A large portion of data, on an average 113 Gbits$/$day will be downloaded at ISSDC obtained using the four detectors of VELC on board ADITYA-L1. The data will consist of coronal images in continuum and spectra in three emission lines. These data need to be separated as per different observing programs and analysed. Number of software programs have been developed using similar space and ground base data. The sample images obtained in the laboratory with the detectors of VELC payload with uniform light have been used to test the programs and calibrate the detectors performance.   
The raw data from space are generally compressed in binary format. The software programs involve to check the correctness of data transmitted, converting into a usable format for analysis by generating the data at various levels. There are number of corrections to be made to the observed images which will be done in steps to verify the image processing algorithms and to confirm the correctness of procedure adopted. Also some information, such as satellite pointing coordinates, time of observation etc., about the images has to be added in the header of the image from other set of downloaded data in separate files. Different scientists may need the data corrected at different stages. Some may need decompressed data with header information only. Others may need data corrected for dark, flat-field, geometrical correction etc. Therefore, data are generated in L0, L1, L2 format with different stages of corrections to meet the requirements of all the scientists. 
Hence, data pipeline software has been developed in Linux OS with IDL as the programming language. The design is based on modular functions, which are processed serially. Each module is sequentially run with the output of one fed to the next module. The overall scheme with different modules is presented below.
 An automated IDL-based Image processing software for the VELC data has been planned. As VELC has four detectors, at POC, the data pipeline software in 4 parallel processes (one process for each detector) will be used. The processing takes place serially for each parallel operation, with the output of one module being the input to the subsequent one. 

\subsection{\textbf{Quick Look Display of VELC data}}

\begin{figure}[h]
\centerline{\includegraphics[width=3.4in, scale=0.5]{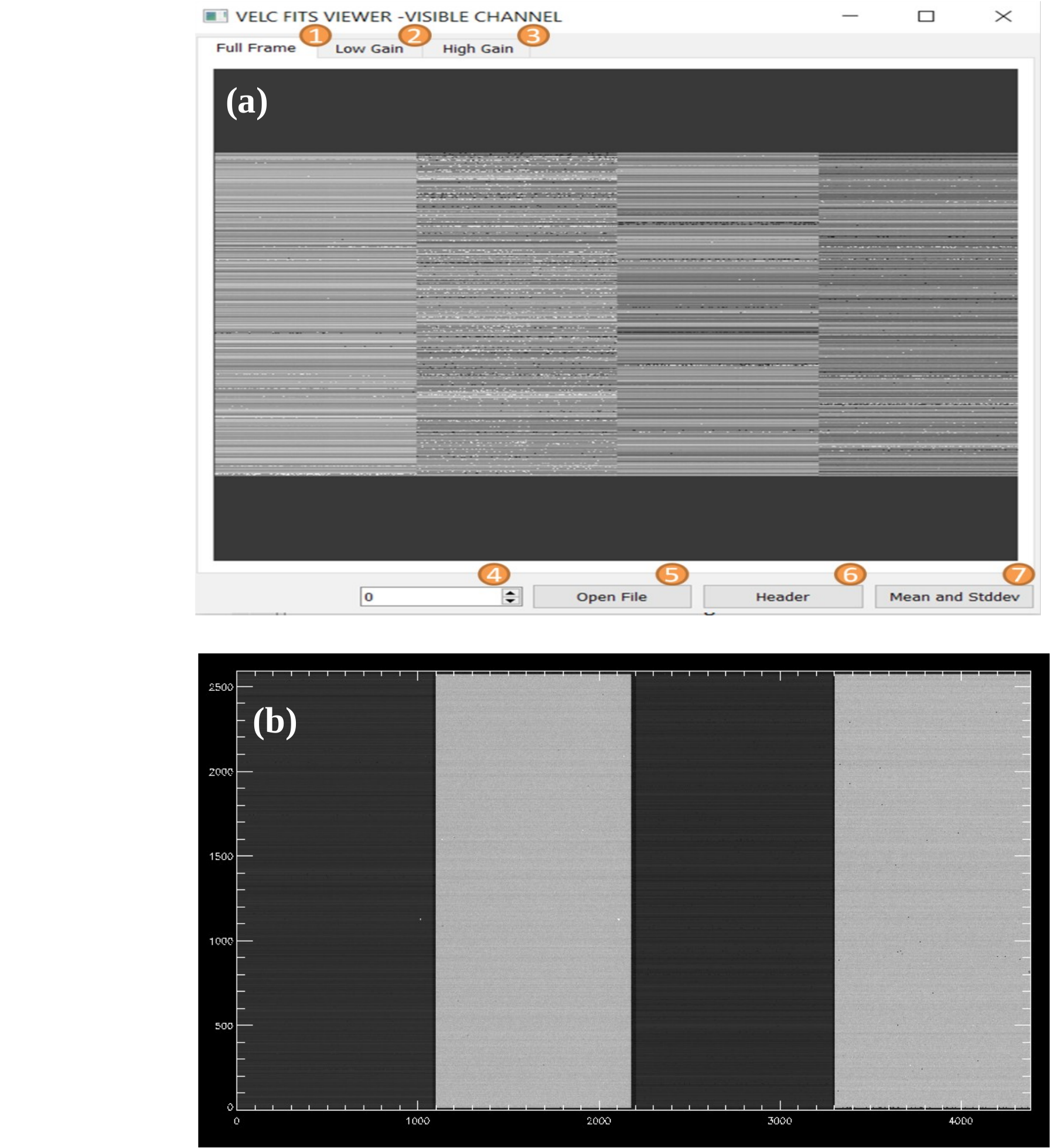}}
\caption{The figure (a) shows the displayed images on the computer screen obtained using CMOS detector without any light source (Imaging $/$ spectroscopic mode). Horizontal lines seen represent rows. Four distinct parts of the image are seen in the vertical direction. Left-side vertical portion of the image represents top-port low gain data, 2nd part shows top-port high gain data, 3rd part represents bottom-port low gain and right-side portion shows the bottom-port high gain data. Figure (b) shows the same taken with light source. }
\label{fig:4}
\end{figure}

A real-time quick look software to verify the health check of VELC raw data that will display the LIVE image of continuum channel and three spectroscopy channels as shown in Figure \ref{fig:4} has been developed. Top panel (marked ``a'') of the figure shows the dark image obtained with CMOS detector in the low and high gain, simultaneously. Bottom panel marked ``b'' shows the images taken with uniform light source. 

The language used is python based GUI Software to display live Quick look data from VELC, compute mean and Standard Deviation (SD) of images. The downloaded raw data is in compressed binary format having two ports - Top and bottom ports as detector has two readouts. Two images taken at same time are packed in single file, one with low gain and other with high gain data. The fits viewer software displays the images obtained using CMOS detector having 2 read out ports (top and bottom) and 2 gains  (low and high gain), separately, as seen in Figure \ref{fig:4}. The display of spectrum taken with IR detector is simple image because of single readout and one gain (high or low gain data). 

Various tabs shown in figure \ref{fig:4}a define different functions to see the parameters of the image.  To display the image on the screen one can choose the file using tab no. 5. Then one can, click on the up arrow to display the next frame in the folder or click on the down arrow for the previous frame. The header tab will print the primary header, secondary header, metadata, telemetry of the file on the console. Click on tab number 7 prints the mean and SD of the image displayed.

\begin{figure}[h]
\centerline{\includegraphics[width=3.4in, scale=0.9]{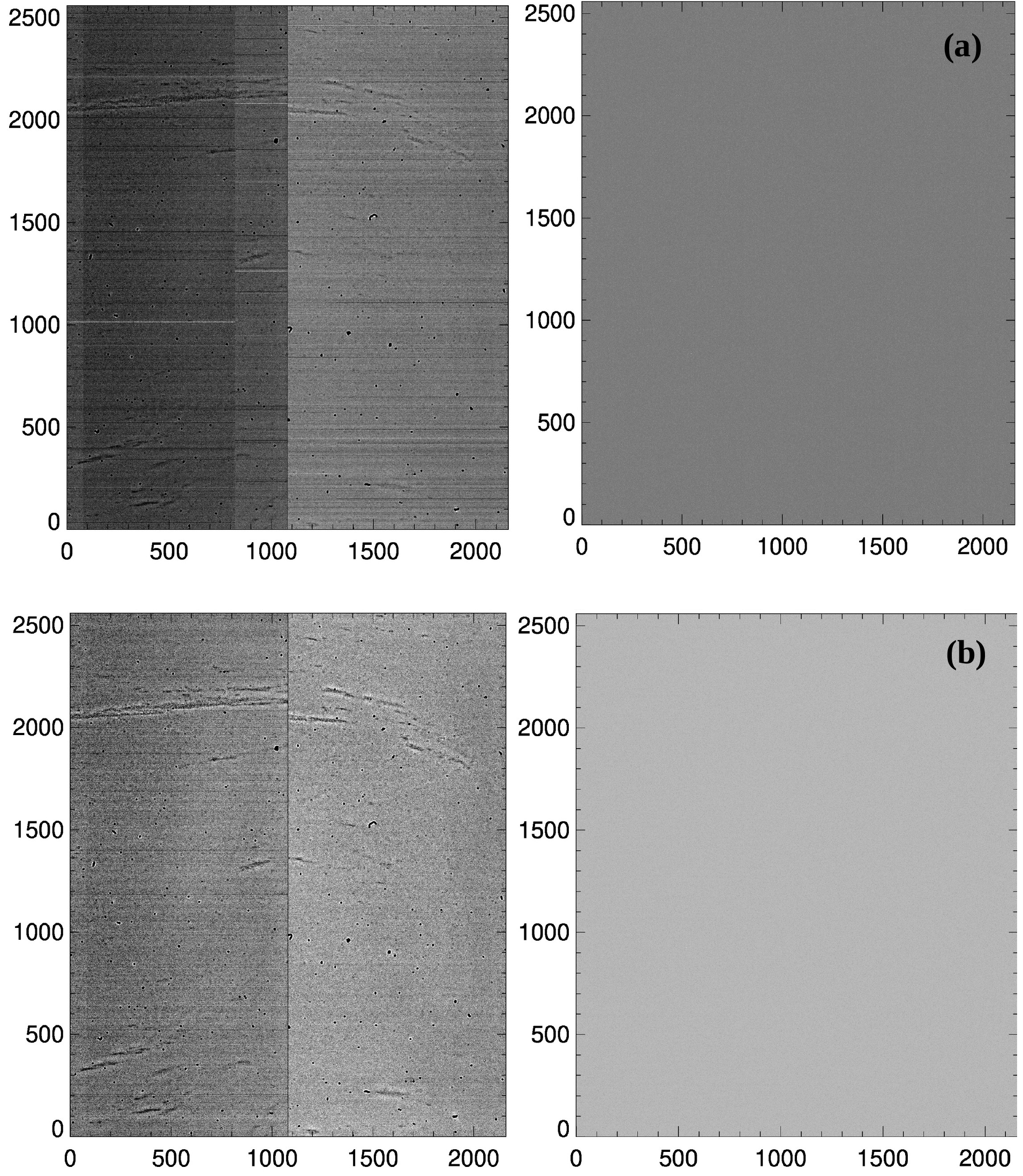}}
\caption{Top left panel of the figure shows the dark corrected (light - dark) image taken with CMOS detector in low gain using uniform light source. Top right panel shows the image after the flat-field correction. Two panels in the bottom row show the images taken in high gain.  }
\label{fig:5}
\end{figure}

\subsection{\textbf{Level-0 Data }}
Images obtained, image information and other related information are saved in different files onboard. These are transmitted to the ground in separate files along with the compressed data in binary format. L0-level data are generated using downloaded images from satellite  and related auxiliary files. Level-0 data contain decompressed binary data of the images  with header having information about type of observations, exposure time, epoch of observations etc. SPICE stands for Spacecraft Planet Instrument Camera-Coefficients and Events. SPICE is a package that organizes the aux data in a collection of useful, called ``SPICE kernels". Each spice kernel file give different parameter of payload and spacecraft. Detailed information about SPICE Kernels, TCT (Time Correlation Table), HK (House Keeping data from payload) will become known in the PV phase. 

\subsection{\textbf{Level-0 to Level-1}}

\begin{figure}[ht!]
\centerline{\includegraphics[width=6.4in, scale=0.5]{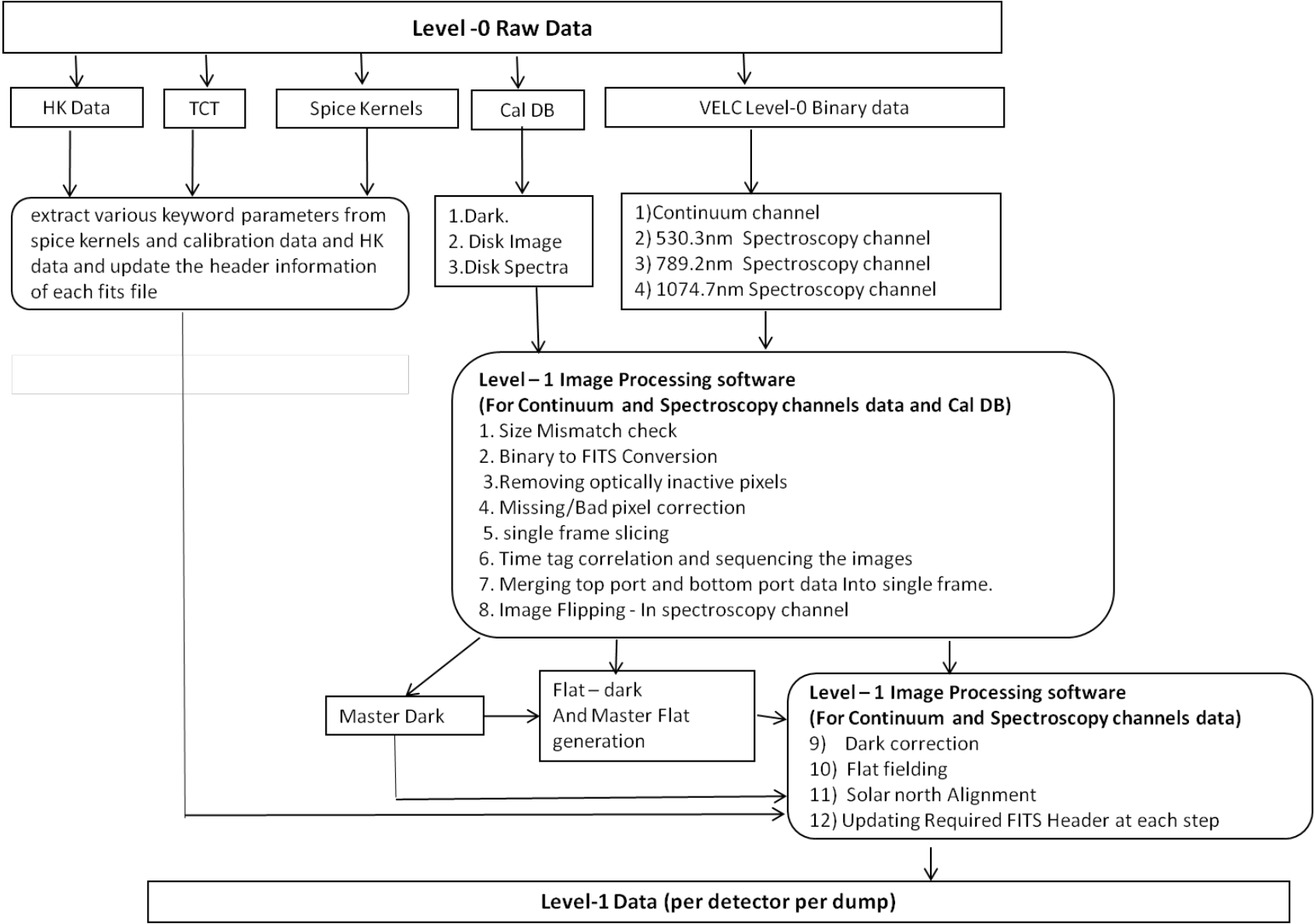}}
\caption{The figure shows the sequence of operations to convert L0 to L1 for VELC data. }
\label{fig:6}
\end{figure}

The data from VELC will be clubbed together for every half an hour and saved onboard in the memory as a ``Dump''. These dumps will be downloaded when satellite comes in the line-of-sight of the download station.  VELC IDL-based image processing software takes the L0 as input and automatically extracts the data from each dump. Then the process separates the dump data into four parallel chains depending of the file naming convention, one for each detector. In addition, the software will extract various keyword parameters from spice kernels, calibration data, and HK data to update the header information of fits file of each image. The L0 calibration data will also be input to the L1 software processing folder,  required for correcting data. Figure \ref{fig:6} shows the sequence of operations done in this process.

It may be noted that the L0 data includes decompressed raw binary data of the images of solar corona and spectra taken with VELC detectors and additional files to update the header information of the image. The decompressed binary data will be first checked for mismatch check, if the size of the file is same as expected. Then, binary to FITS conversion and other corrections such as removing optically inactive pixels, dead$/$less active$/$hot pixel, image slicing in 4 parts due to two ports and two gain, merging of top and bottom ports to form two images, one of low gain and other high gain, time tag correlation etc. Additionally, for spectroscopy channels, the images need to be corrected for flipping caused due to VELC optics. Using the calibration data obtained onboard, master dark will be  generated to subtract dark signal from the coronal images and spectra. Using the calibration data obtained onboard, master dark will be generated to subtract dark signal from the coronal images and spectra.  This calibration data is planned to be taken every three months.  Then, images will be flat fielded and aligned using satellite pointing information. Top left panel of Figure \ref{fig:5} shows the dark corrected (light - dark) image taken with sCMOS detector in low gain with uniform light source. The horizontal lines in the figure are along the row and due to difference in amplification at the readout process. Apart from these horizontal lines there are some random patterns may be due to chip. Top right panel of the figure show the image corrected for the flat-field (pixel to pixel variations due to different response of the pixel to the light). The figure indicates that the image has become uniform free from the horizontal lines and patterns as expected because of uniform light used for taking images. Two panels in the bottom row show images taken in high gain after correction for the dark and flat-field signal.

\subsection{\textbf{Level-1 to Level-2}}

\begin{figure}[ht!]
\centerline{\includegraphics[width=6.4in, scale=0.9]{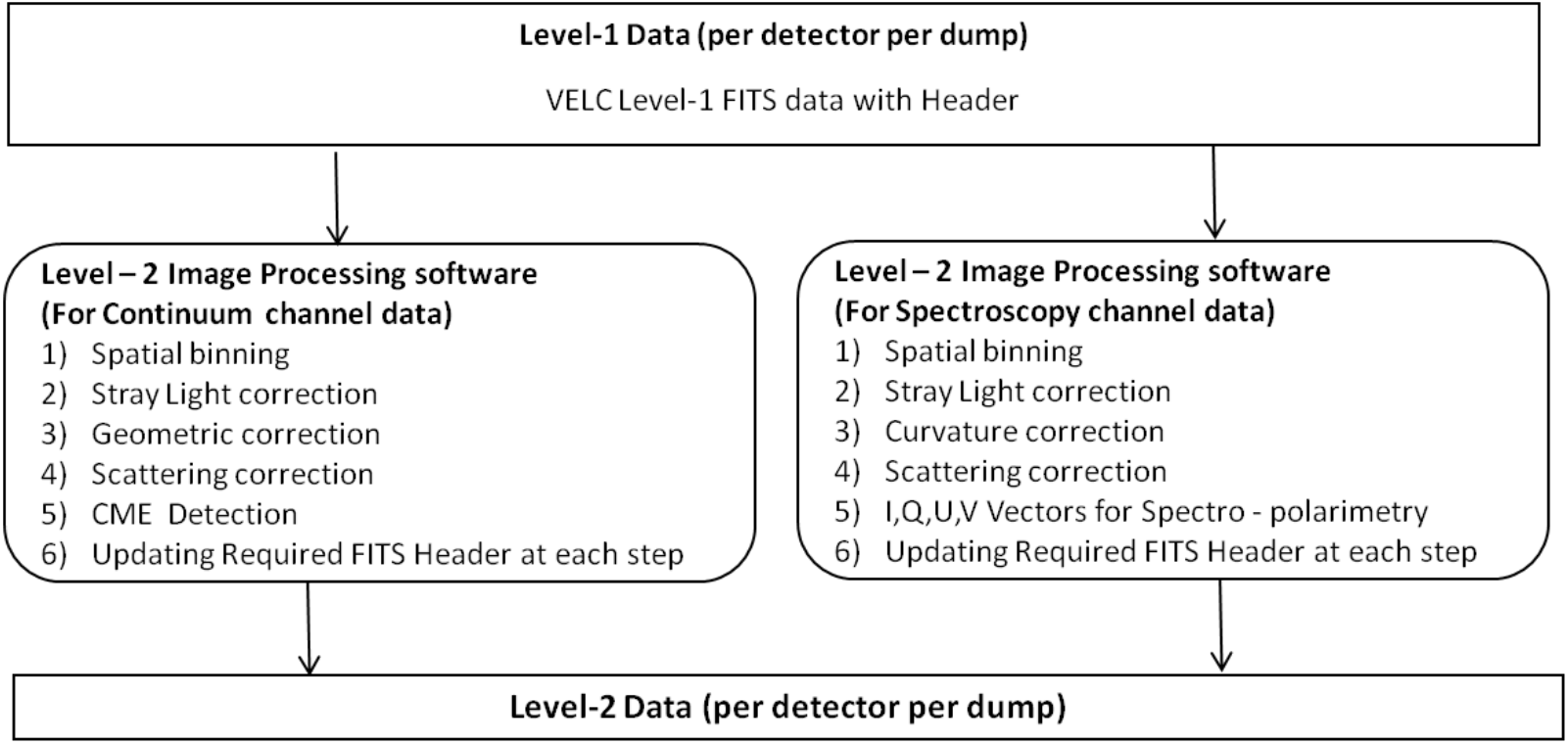}}
\caption{The figure shows the sequence of operations to convert  L1 to L2 Data pipeline.  }
\label{fig:7}
\end{figure}

L1 output is input to L2 Software, where higher level data processing will be done. After the separation of data according to detector (channel), further analysis of the Level-1 data is planned to be done by 4 separate computers, one for each channels due to some differences in the software for imaging, visible and IR spectroscopic channels. Figure \ref{fig:7} shows the sequence of operations and corrections, such as spatial binning, stray light (pattern due to some unknown reason, if any) , curvature , and scatter (disk scattered light, model dependent),  are planned to be made to the Level-1 data to generate L2 data. After completing all the above mentioned processing of the data, four Level-2 output zip files corresponding to each detector will be generated to be sent to ISSDC. A CME detection and visualization algorithm for the continuum channel is being developed to operate on the downloaded coronal images.

\subsection{\textbf{Offline Quick Look Display and Public outreach product }}
Image enhancement to the corrected L1 fits images to generate JPG/PNG images and videos for the selected interesting events for public outreach and to be sent ISSDC is planned. 1) JPG Images of coronal Images and spectra, 2) Video of detected CME in the continuum channel, 3) Video of spectra showing variation with time made from the data taken in ``Sit and Stare mode'' observations 4) Coronal images in emission line constructed from ``Raster scan'' observations, will be generated after the analysis and calibration of data obtained for the public view first and later for the scientific results after the confirmation and detailed calibration of the data.

\section{Summary}
 The data pipeline and software to analyse large volume of data in binary format from VELC, automatically and generate fits images for further processing have been developed. The requirement of different scientific studies, observational parameters for VELC listed in ``Proposal submission form'' have been discussed in various scientific forms and ISRO review committees. The software developed has been tested on data similar to those expected from VELC. After the ADITYA-L1 launch, the performance evaluation of the VELC data pipeline will be done and modified, if needed.

\section{Acknowledgements }
Authors thank all the Scientists$/$Engineers at the various centres of Indian Space Research Organisation (ISRO) such as URSC, LEOS, SAC, SAC, ISTRAC, VSSC etc. and Indian Institute of Astrophysics who have made great contributions to the mission to reach at the present state. Authors gratefully acknowledge the financial support from ISRO for this project. The authors thank the referee for the useful comments.

\end{document}